\documentclass[aps,showpacs,pra,twocolumn,superscriptaddress]{revtex4}
\usepackage{graphicx}
\usepackage{amssymb}
\usepackage{epstopdf}
\usepackage{color, amsmath, bm}
\usepackage[hidelinks]{hyperref}
\newcommand{\be}{\begin{equation}}
\newcommand{\ee}{  \end{equation}}
\newcommand{\ba}{\begin{eqnarray}}
\newcommand{\ea}{  \end{eqnarray}}

\begin{document}

\title{Geometric phase accumulated in a driven quantum system coupled to a structured environment}
\author{Paula I. Villar}
\affiliation{Departamento de F\'\i sica {\it Juan Jos\'e
Giambiagi}, FCEyN UBA and IFIBA CONICET-UBA, Facultad de Ciencias Exactas y Naturales,
Ciudad Universitaria, Pabell\' on I, 1428 Buenos Aires, Argentina.}
\author{Alejandro Soba}
\affiliation{ Centro  At\'omico  Constituyentes,  Comisi\'on  Nacional  de  Energ\'\i a  At\'omica,
Avenida  General  Paz  1499,  San  Mart\'\i n,  Argentina}
\date{\today}                                           

\begin{abstract}

We study the role of driving in a two-level system evolving under the presence of a structured environment
in different regimes. 
We find that adding a periodical modulation 
to the two-level system can greatly enhance the survival of the geometric phase for many time periods in an intermediate coupling to the environment.  In this regime, where there are some non markovian features characterizing the dynamics, we have noted that adding driving to the system leads to a suppression of non-markovianity altogether, allowing for a smooth dynamical evolution and an enhancement of the robustness condition of the
geometric phase.
As the model studied herein is the one used to model experimental situations such as hybrid quantum classical systems feasible with current technologies, we positive believe this knowledge can aid the search for physical set-ups that best 
 retain quantum properties under dissipative dynamics.

\end{abstract}

\maketitle

The state of a point like discrete energy level quantum
system  interacting with a quantum field acquires a geometric phase (GP) that is independent of the state of the
field \cite{berry}. The phase depends only on the system's path
in parameter space, particularly the flux of some gauge field enclosed by that path. 
Due to its topological properties and close connection with gauge theories of quantum fields, 
the GP has recently become a fruitful venue of investigation to infer features of the quantum system.
For pure field states, the GP is said to encode information about the number of particles in the  field \cite{caridi}. 
If the field is in a thermal state, the GP encodes information about its temperature, and so it has been used in a proposal to measure the 
Unruh effect at low accelerations \cite{martinez1}. Furthermore, in \cite{martinez2}, it has been proposed as a high precision thermometer in order to infer the 
temperature of two atoms interacting with a known hot source and an
unknown temperature cold cavity. In this context, the study of the GP in open quantum systems has been a subject of investigation lately.
The definition of the geometric phase for non-unitary evolution was first stated in \cite{Tong}.
This definition has been used to measure the corrections of the GP in a non-unitary evolution \cite{prl} and to explain the
noise effects in the observation of the GP in a superconducting qubit \cite{leek,pra}.
The geometric phase of a two-level system under the influence of an external environment has been studied in a wide variety of scenarios \cite{papers}.
It has further been used to track traces of quantum friction in an experimentally viable scheme of a neutral particle traveling at constant velocity in front of a dielectric plate \cite{nature} and in a very simplistic analytical model of an atom coupled to a scalar quantum field \cite{epl}. 

The coupling of the quantum system to the environment
is described by the spectral density function. If the system
couples to all modes of the environment in an equal way the spectrum of the reservoir is flat. If, otherwise,
the spectral density function strongly varies with the frequency of the environmental oscillators, the environment
is said to be structured. In this type of environment the memory effects induce a feedback of information from the environment into the system. They are therefore called non-markovian \cite{breuer}.
Numerous works have investigated the presence of non-markovianity in a variety of scenarios in quantum open systems so as to determine
whether non-markovianity is a useful resource for quantum technologies. 
It  has been studied how the presence of a driving field affects the non-markovian features of a quantum open system.  
For instance, studies which assessed the effectiveness
of optimal control methods \cite{Zhu,Krotov} in open quantum system
evolutions showed that non-markovianity allowed for an improved
controllability \cite{Schmidt, Reich, Triana}. Likewise, the non-markovian effects were associated to the reduction of efficiency in dynamical
decoupling schemes \cite{addis} and accounted for corrections to the GP acquired \cite{pra1,Luo,Oh}.

In this work, we investigate to what extent external
driving acting solely on the system can increase 
non-markovianity (and therefore modify the geometric phase) 
with respect to the undriven case. To this end, we
consider a two-level system described by a time-periodic
Hamiltonian interacting with a structured environment.
It has been recently shown that the driving has a peculiar effect on the non-markovian character of the system dynamics: it can generate
a large enhancement of the degree of non-markovianity with respect
to the static case for a weak coupling between the system and environment \cite{Poggi}. 
The importance of the driven two-state model is especially pronounced in quantum computation and quantum technologies, where one or more driven qubits constitute the basic building block of quantum logic gates \cite{nielsen}.
Geometric quantum computation exploits GPs to implement universal sets of one-qubit and two-qubit gates, whose realization finds versatile platforms in systems of trapped atoms \cite{Duan}, quantum dots \cite{solinas} and superconducting circuit-QED \cite{Faoro}.  
Different implementations of qubits for quantum logic gates are subjected to different types of environmental noise, i.e., to different environmental spectra. Since the model studied herein can be implemented in these experimental contexts, using real or artificial atoms, it is important to unveil the time behavior of the qubit geometric phase for driven systems. 
We shall only focus on weak or intermediate coupling since we try to track traces of the geometric phase, which is literally destroyed under a strong influence of the environment. 
This means that while there are non-markovian effects  that induce a correction to the unitary GP, the system maintains its purity for
 several cycles, which allows the GP to be observed. It is important to note that if the noise effects
induced on the system are of considerable magnitude, the coherence terms of the quantum system  are rapidly destroyed and the GP literally disappears \cite{papers}.
This knowledge  can aid the search for physical set-ups that best retain quantum properties under dissipative dynamics.

This paper is structured as follows. In Sec. \ref{modelo} we present the model consisting of a two-level system described by a time-periodic Hamiltonian interacting with a structured environment.  In Sec. \ref{dinamica}, we numerically solve the dynamics of the system for different regimes  through the hierarchy method beyond the rotating wave approximation. In Sec. \ref{fase} we compute the geometric phase
for a two-level driven system  and analyze its deviation from the unitary geometric phase under  different regimes.  Since we want  to track traces of the geometric phase, which is literally destroyed under a strong influence of the environment, we shall restrict our study to two situations: (A) weakly coupling  and (B) intermediate coupling. Therein, we analyze the robustness condition of the geometric phase acquired by the driven two level system and the best scenarios for its experimental detection. 
Finally, in Sec. \ref{conclusiones}, we summarize the results and present conclusions.

\section{The Model} \label{modelo}
We shall consider a two-level system described by a time-periodic Hamiltonian interacting with an environment.
  The total Hamiltonian which describes this model reads (we set $\hbar=1$ from here on)
\begin{equation}
H=  \bar{\omega}_0 (t) \sigma_+\sigma_- + \sigma_x \sum_k  (g_k b_k + g_k^*b_k^{\dagger}) + \sum_k \bar{\omega}_k b_k^{\dagger} b_k,
\end{equation}
where  $\sigma_{\pm}= \sigma_x \pm i \sigma_y$ (with $\sigma_{\alpha}$ ($\alpha=x,y,z$) the Pauli matrices) and $b_k$, $b_k^{\dagger}$ the annihilation and creation operators corresponding to the $k-$th mode of the bath. The coupling constant is $g_k$ and $\bar{\omega}_0(t)$ is the time-dependent energy difference between the states $|0\rangle$ and $|1\rangle$ of the two-level system. We shall assume it has the following form:
\begin{equation}
\bar{\omega}_0(t)= \bar{\Omega} + \bar{\Delta} \cos({\bar {\omega}_D} t).
\end{equation}
The exact dynamics of the system in the interaction picture has been derived in \cite{Tanimura}.
If the qubit and the bath are initially in a separable state, i.e. $\rho(0)=\rho_s(0)\otimes \rho_B$, the formal solution is:
\begin{eqnarray}
\tilde{\rho}_S(t) &=& {\cal T} \exp\bigg(-\int_0^t dt_2\int_0^{t_2} dt_1 \tilde{\sigma_x}^\times (t_2) \\
&& [C^R(t_2-t_1)\tilde{\sigma_x}^{\times}(t_1) + i C^I(t_2-t_1)\tilde{\sigma_x}^{\circ}(t_1)] \bigg), \nonumber
\label{rhoexacta}
\end{eqnarray}
where  ${\cal T}$ implies the chronological time-ordering operator and  $\tilde {o}$, denotes the expression of the operator $o$ in the interaction picture. We have further introduced the following notation $A ^{\times} B=[A,B]= AB-BA$ and $A^{\circ}B= \{A,B\}= A B+ B A$.
$C^R(t_2-t_1)$ and $C^I(t_2-t_1)$ are the real and imaginary parts of the bath time-correlation function, defined as
\begin{eqnarray}
C(t_2-t_1) &\equiv& \langle B(t_2) B(t_1) \rangle = {\rm Tr}[B(t_2)B(t_1)\rho_B] \nonumber \\
&=& \int_0^{\infty} d\omega J(\omega) e^{-i \omega (t_2-t_1)}
\end{eqnarray} 
and 
\begin{equation}
B(t)=\sum_k \bigg(g_k b_k \exp(-i \omega_k t) + g_k^*b_k^{\dagger} \exp(i \omega_k t)\bigg). \nonumber 
\end{equation}
Eq.(\ref{rhoexacta}) is difficult to solve directly. An effective method for obtaining a solution has been developed by defining a set of hierarchy equations  \cite{Tanimura,Tanimura2,Sun}.
The key condition in deriving the hierarchy equations is that the correlation function can be decomposed into a sum of exponential functions of time.
At finite temperatures, the system-bath coupling can be described by the Drude spectrum, however, if we consider qubit devices, they are generally prepared 
in nearly zero temperatures. Then we shall consider a Lorentz type spectral density $J(\omega)$,

\begin{equation}
J(\omega)= \frac{{\bar{\gamma}_0}}{2 \pi} \frac{\lambda^2}{(\omega- \bar{\Omega})^2 + \lambda^2},
\end{equation}
and the hierarchy method can also be applied \cite{Sun2}. As has been stated in \cite{Poggi}, this method can be used if i) the initial state of the system plus bath is separable, ii) the interaction Hamiltonian is bilinear, and iii) if the environmental correlation function can be cast in multi-exponential form.
In this case, $\bar{\gamma_0}$ is the coupling strength between the system and the bath and 
$\lambda$ characterizes the broadening of the spectral peak, which is connected to the bath correlation time $\tau_c= \lambda^{-1}$. The relaxation 
time scale at which the state of the system changes is determined by $\tau_r=\bar{\gamma_0}^{-1}$.
At zero-temperature, if we consider the bath in a vacuum state, the correlation function can be expressed as
\begin{equation}
C(t_2-t_1)= \frac{\lambda \bar{\gamma_0}}{2} \exp([-(\lambda + i \bar{\Omega})|t_2-t_1|])
\end{equation}
which is the exponential form required for the hierarchy method. The advantage of solving the dynamics of the system by this method is that we
can take an insight of different regimes of the dynamics. For example, in the limiting case $\bar{\gamma_0} \ll \lambda$, i.e. $\tau_c \ll \tau_r$, we have a flat
spectrum and the correlation tends to $C(t_2-t_1)\rightarrow \bar{\gamma_0} \delta (t_2-t_1)$. This is the so called markovian limit. Therefore, we can study the
full spectrum of behavior by solving the hierarchy method, which can be expressed as 
\begin{widetext}
\begin{equation}
\frac{d}{d\tau}\rho_{\vec{n}}(\tau)= -(i H_s[\tau]^{\times}+ \vec{n}.\vec{\nu})\rho_{\vec{n}}(\tau) - i \sum_{k=1}^2 \sigma_x^{\times}\rho_{\vec{n}+\vec{e}_k}(\tau)- i\frac{{\gamma_0}}{2}
\sum_{k=1}^2 n_k [\sigma_x^{\times} + (-1)^k \sigma_x^{\circ}] \rho_{\vec{n}-\vec{e}_k}(\tau),
\label{hierarchy}
\end{equation}
\end{widetext}
where we have defined dimensionless parameters variables $\tau=\lambda t$ and $x=\bar{x}/\lambda$ where $x$ is any parameter with units of energy in the model described.
The subscript $\vec{n}=(n_1,n_2)$ with integers numbers $n_{1(2)} \geq 0$, and $\rho_S(t) \equiv \rho_{(0,0)} (t)$. This means that the ``physical" solution
is encoded in $\rho_{(0,0)} (t)$ and all other $\rho_{\vec{n}}(\tau)$ with $\vec{n} \neq (0,0)$ are auxiliary operators implemented for the sake of computation. We have defined the vector $\vec{\nu}=(\nu_1,\nu_2)=(1-i \Omega, 1 + i \Omega)$.
The hierarchy equations are a set of linear differential equations, that can be solved by using a Runge Kutta rutine. For numerical computations, the hierarchy equations
must be truncated for large $\vec{n}$. The hierarchy terminator equation is similar to that of Eq.(\ref{hierarchy}) for the term $\vec{N}$, and the corresponding term related to $\rho_{\vec{N}+\vec{e}_k}$ is dropped \cite{Tanimura}. The numerical results in this paper have been all tested and converged, using a maximum value of $\vec{N}=(25,25)$.
We shall take advantage of this model, whose non-markovian properties has been studied in \cite{Poggi}, and set the scenario to study the corrections to the GP for a driven two-level system.

\section{Environmentally induced Dynamics} \label{dinamica}

We begin by studying the environmentally induced dynamics by considering a qubit with no driving at all ($\Delta=0$). In this case, we must consider a qubit and a dipolar coupling to the cavity mode, for example. This means, that the dynamics of the system contemplates decoherence and dissipation as well as variation of the population numbers (in contrast to the spin boson model). The density matrix for this case has a formal expression as
\begin{equation}
\rho_s(\tau)= \bigg( 
\begin{matrix}
\rho_{11}|G(\tau)|^2 & \rho_{12} G(\tau) \\
 \rho_{21} G^*(\tau) & 1-\rho_{11}|G(\tau)|^2
\end{matrix}
\label{rho}
\bigg)
\end{equation}
where $G(\tau)$ is a single-complex valued function that characterizes the dynamics of the system. We herein do not write its explicit form since
we shall solve the problem numerically through the hierarchy approach.

\begin{figure}[h!]
	\includegraphics[width=9.5cm]{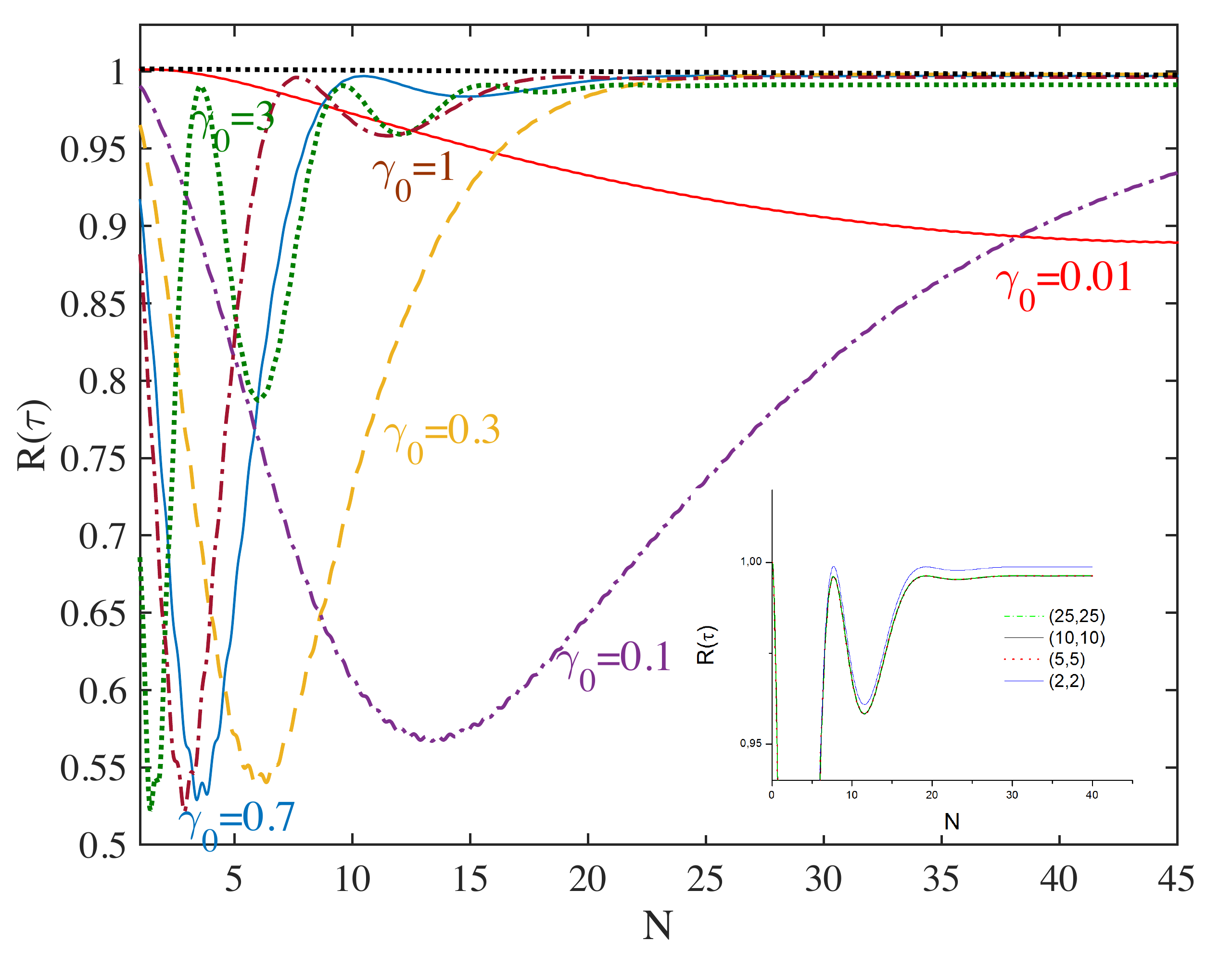}
 	\caption{(Color online) We plot the loss of the quantum state purity $R(\tau)$ as function of time $N$($\omega_0 \tau= N 2\pi $ number of cycles). We can see that as the coupling constant with the bath increases for a fixed value of $\lambda=0.01$ ($\gamma_0=\bar{\gamma}_0/\lambda$),  the dynamical behavior is modified.  Orange dashed line is $\gamma_0=0.3$ and dot dashed purple line for $\gamma_0=0.1$ represent situations of $\tau_r > \tau_c$. Dot-dashed brown line is for $\gamma_0=1$, blue solid line for $\gamma_0=0.7$ are situations of $\tau_r \sim \tau_c$. In the inset we show different solutions for  $\gamma_0=0.1$ by varying the order of truncation. We can state that from $\vec{N}=(10,10)$ we can obtain a converged positive reduced matrix $\rho(\tau)$.	Parameters used: $\Delta=0$, $\vec{N}=(25,25)$, $\Omega=20$.}
 	\label{Fig1}
 \end{figure}

Decoherence time $\tau_D$ is mostly known as the timescale at which the quantum interferences are suppressed. This is formally true for a purely dephasing process where noise only affects the off-diagonal terms of the reduced density matrix. However, Eq.(\ref{rho}) describes a process where populations and off-diagonal terms are both affected by the presence of noise.
Qualitatively, decoherence can be thought of as the deviation of probabilities measurements from the ideal intended outcome.
Therefore decoherence can be understood as fluctuations in the Bloch vector  $\vec{R}$ induced by noise. 
In a wider sense, we will represent decoherence as the change of $|\vec{R}(\tau)|$ in time, starting from $|R(0)| = 1$ for the initial pure state, and decreasing as long as the quantum state loses purity.  The contributions of the bath to the dynamics of the system, including both dissipation and Lamb shift, are fully contained in the hierarchy equation.
In Fig. \ref{Fig1} we present the absolute value of the Bloch vector of the state system $R(\tau)=\sqrt{x(\tau)^2+y(\tau)^2+z(\tau)^2}$ as a function of time measured in natural cycles $\omega_0 \tau= N 2\pi $ for different values of $\gamma_0$.
In this case, we can note that the trajectory differs substantially from the unitary one, meaning the system's dynamics is affected by the noise effects. In the case the unitary dynamics is considered, $\gamma_0=0$ and $R=1$ for all times. 

We can notice that the dynamical behavior is modified as  the coupling constant $\gamma_0$ is increased.  It is interesting to see the interplay between time and $\gamma_0$: a stronger bath can initially produce less damage on the dynamics but has a stronger effect in the renormalization of the frequency. A weak-coupling has a more ``adiabatic" modification of the dynamics in an equal period of time. In Fig. \ref{Fig1}, we have set $\lambda$ fixed. As $\gamma_0$  increases, the relaxation time $\tau_r$ of the system decreases and $\tau_r \sim \tau_c$. The presence of oscillations in the Bloch vector $R(\tau)$ for short times, as $\gamma_0$ becomes similar to $\lambda$ indicates non markovian dynamics induced by the reservoir memory and describing the feedback of information and/or energy from the reservoir into the system \cite{addis}. We can see that as long as $\bar{\gamma_0}/\lambda < 1$, the systems exhibits a markovian dynamics (orange-dashed line, purple dot-dashed line and red solid line). On the other side, if $\bar{\gamma_0}/\lambda \geq 1$, there are non markovian features in the system's dynamics. 
We can notice that as long as $\bar{\gamma_0} \leq \lambda$ and $\lambda \ll 1$, the behavior remains similar to that of $\lambda \rightarrow 0$ (markovian since $\tau_c \rightarrow \infty$). However, as $\lambda$ increases, the environmentally-induced dynamics is considerably modified, introducing oscillations again. So, with this kind of environment we can simulate different regimes by the solely selection of the $\bar{\gamma_0}$ and $\lambda$ parameters.  In the inset of  Fig. \ref{Fig1}, we show a simulation for different truncations of the system of equations for $\gamma_0=1$. We show that by setting the order of truncation in 25, we already obtain  a converged positive reduced matrix $\rho(\tau)$.

 \begin{figure}[h!]
 	\includegraphics[width=9.3cm]{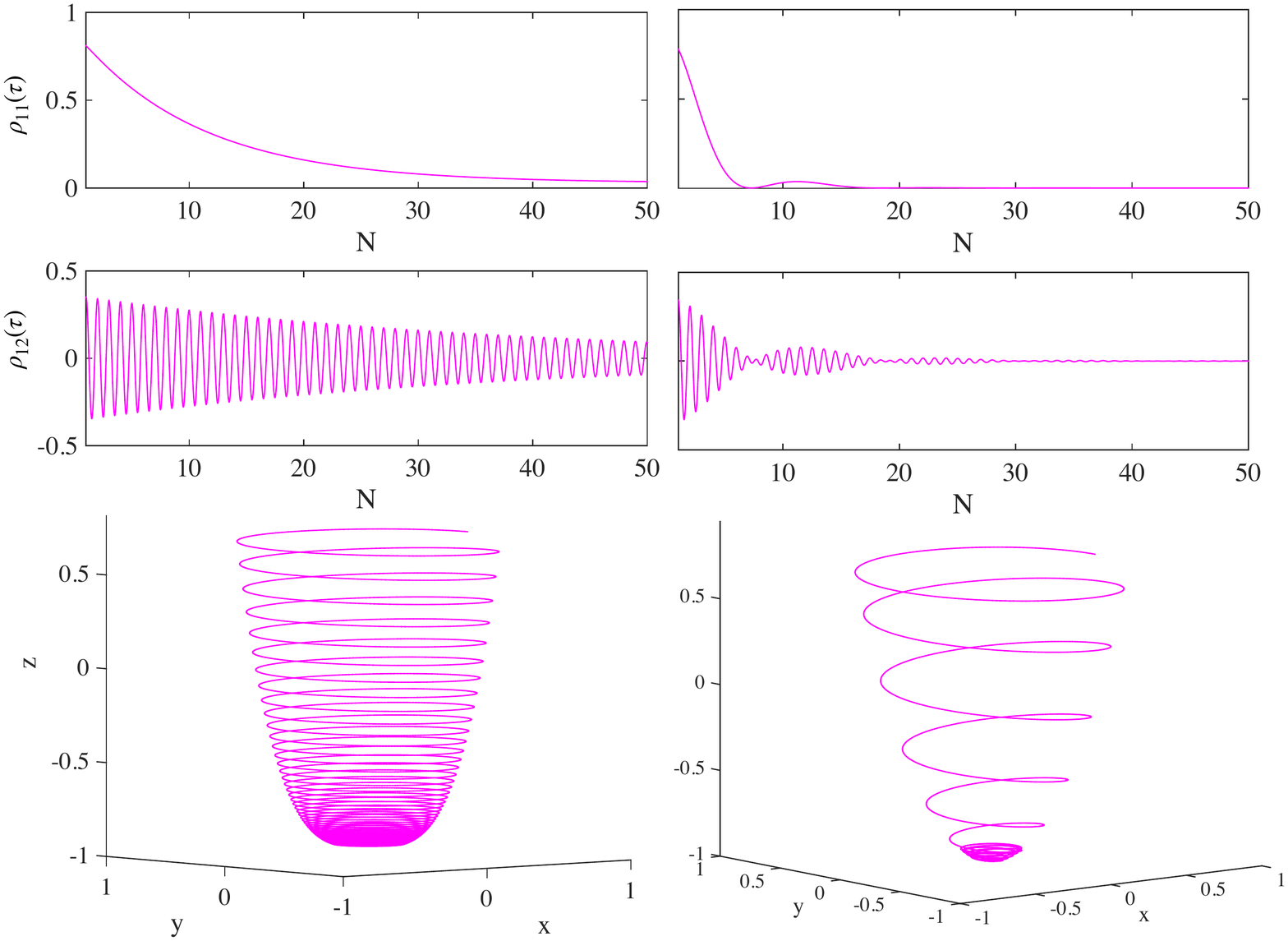}
 	\caption{(Color online) We plot different dynamics for a fixed value of  $\lambda$, but different values of the $\bar{\gamma}_0$ parameter. The left column is for $\bar{\gamma_0}/\lambda <1$ and the right for $\bar{\gamma_0}/ \lambda \geq 1$. On top we show the behavior of $\rho_{11}(t)$ and the absolute value of $\rho_{12}(t)$ in each case.  We can see some time revivals on the right reduced matrix elements. The lower plots show the trajectory ($\vec{R}=(x,y,z)$) of the two level system in the Bloch sphere.  Parameters used: $\Delta=0$, $\vec{n}=(25,25)$, $\Omega=20$, $\tau_c=100$. }
 	\label{Fig3}
 \end{figure}

In Fig. \ref{Fig3}, we compare the dynamics of two different environmental situations: the left column is for $\bar{\gamma_0}/\lambda <1$, and the right one for $\bar{\gamma_0}/\lambda \geq 1$,  both evolutions are simulated for fixed $\bar{\Omega}$ and zero driving ($\Delta=0$). In this example, we can see that when $\tau_c < \tau_r$, the system presents a markovian evolution. On the other hand, if $\tau_c > \tau_r$, non-markovian effects can be seen, for example by accelerating the transition between quantum states and  revivals for longer times. For initial short times, the spontaneous decay of the atom can not only be suppressed  or enhanced, but also partly reversed, when non-markovian oscillations induced by reservoir memory effects are present. As has been shown, by choosing the right set of parameters, we can simulate different type of environments and obtain the corresponding dynamics beyond the rotating wave approximation.

\section{Correction to the Geometric Phase}\label{fase}

In this section, we shall compute the geometric phase
for the central spin and analyze its deviation from
the unitary geometric phase for a two-level driven system.
A proper generalization of the geometric phase for
unitary evolution to a non-unitary evolution
is crucial for practical implementations of geometric
quantum computation. In \cite{Tong}, a quantum kinematic
approach was proposed and the geometric phase 
(GP) for a mixed state
under non-unitary evolution has been defined as
\begin{eqnarray} 
\Phi & = &
{\rm arg}\{\sum_k \sqrt{ \varepsilon_k (0) \varepsilon_k (T)}
\langle\Psi_k(0)|\Psi_k(T)\rangle \nonumber \\ 
& & \times e^{-\int_0^{T} dt \langle\Psi_k|
\frac{\partial}{\partial t}| {\Psi_k}\rangle}\}, \label{fasegeo}
\end{eqnarray}
where $\varepsilon_k(t)$ are the eigenvalues and
 $|\Psi_k\rangle$ the eigenstates of the reduced density matrix
$\rho_{\rm s}$ (obtained after tracing over the reservoir degrees
 of freedom). In the last definition, $T$ denotes a time
after the total system completes
a cyclic evolution when it is isolated from the environment.
Taking the effect of the environment into account, the system no
longer undergoes a cyclic evolution. However, we will consider a
quasi cyclic path ${\cal P}: T~\epsilon~[0,\tau_S]$ with
$\tau_S=2 \pi/\omega_0$ ($\omega_0$ the system's dimensionless frequency).
When the system is open, the original GP $\phi_u$, i.e. the one that
would have been obtained if the system had been closed, is
modified. This means, in a general case, the phase is
$ \phi_g=\phi_u+ \delta \phi$,
where $\delta \phi$ depends on the kind of environment coupled to
the main system \cite{papers, zanardi}. For a spin-1/2 particle
in $SU(2)$, the unitary GP is known to be $\phi_u= \pi(1+\cos(\theta_0))$.
It is worth noticing that the proposed GP is gauge invariant and leads
to the well known results when the evolution is unitary.

As this method can be used when the initial state of the whole system is separable, we shall
start by assuming $\rho(0)= \rho_s(0) \otimes \rho_{\cal E}(0)$.
The initial state of the quantum system is supposed to be a pure state
 of the form:
 \begin{equation}
|\Psi (0) \rangle=\cos(\theta_0/2) |0 \rangle
+ \sin(\theta_0/2) |1 \rangle .
\nonumber\end{equation}
We shall solve the master equation and then, compute the geometric phase acquired by the quantum system.
If the environment is strong, then the unitary evolution is destroyed in a decoherence time $\tau_D$. Otherwise, we can imagine an scenario where the effect
of the environment is not so drastic. 
In the following, we shall focus on how driving can affect  (or even benefit) the measurement of the geometric phase under different regimes, both weakly coupling and intermediate coupling.  In particular, we shall investigate to what extent external driving acting solely on the system can  correct the geometric phase with respect to the undriven or unitary case.

\subsection{Geometric phase under a weakly coupling}

\begin{figure}[t]
 	\includegraphics[width=9cm]{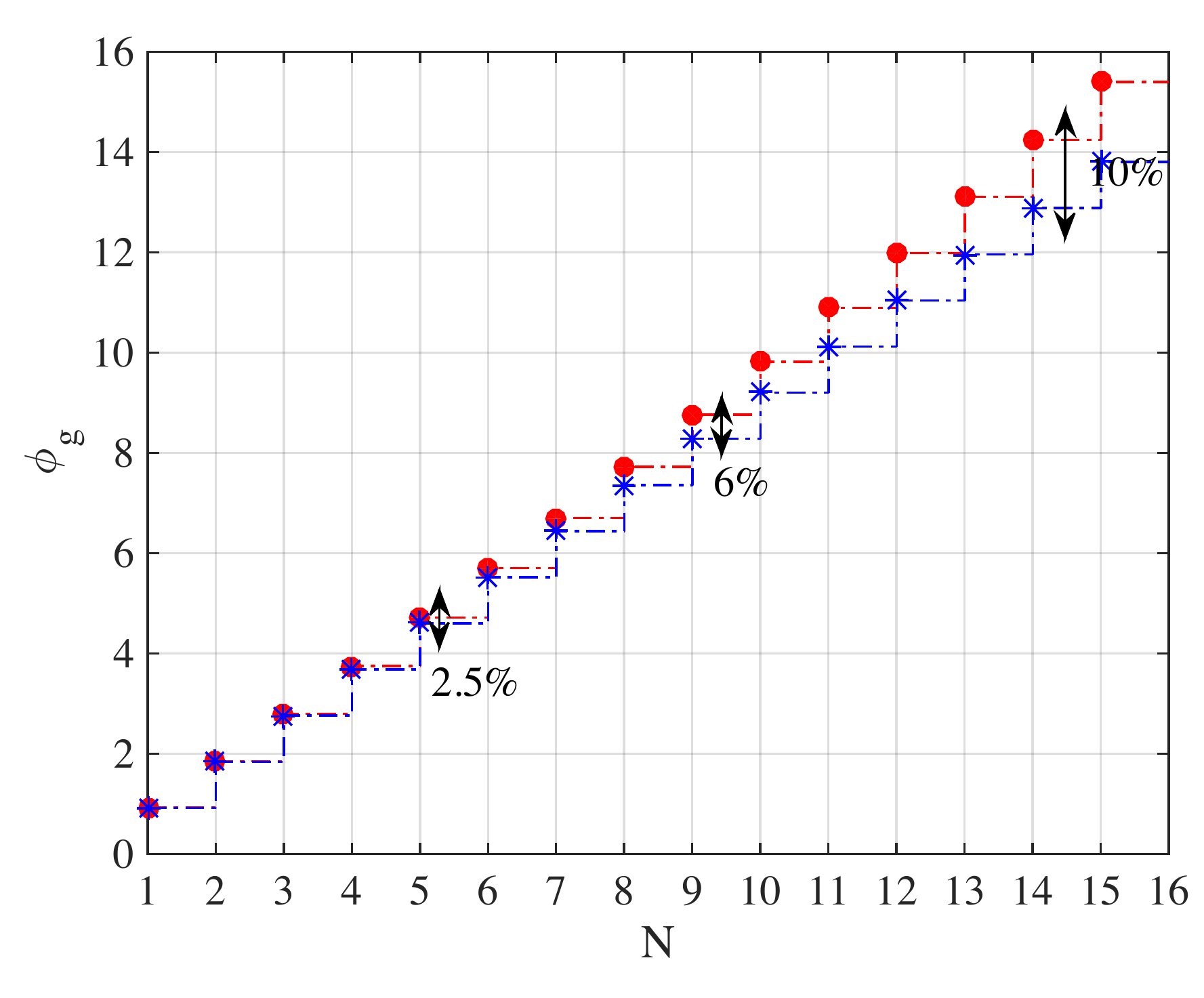}
	\caption{(Color online) Comparison between  the accumulated geometric phase $\phi_g$ for the unitary case (asterisk blue dot-dashed line) and a weak coupled  environment in a markovian regime $\bar{\gamma_0 }\ll  \lambda$ (circled red dot-dashed line) (corresponding to a similar situation to that of the left column of Fig. \ref{Fig3}). Parameters used: $\gamma_0=0.01$, $\Omega=20$, $\theta_0=\pi/4$, $\Delta=0$ and $\omega_D=0$.}
 	\label{Fig4}
 \end{figure}
The dynamics of the driven two-level system comprises of three different dynamical effects, occurring each at a different timescale. Dissipation and decoherence occur at the relaxation timescale $\tau_r$, non-markovian memory effects occur at times shorter or similar to the reservoir correlation timescale $\tau_c$ \cite{addis}. Finally, nonsecular terms cause oscillations in a timescale of the system $\tau_S=(\Omega^2+\Delta^2)^{-1/2}$. Generally, this nonsecular terms can be neglected when $\tau_c \ll \tau_S $. We shall consider the secular regime, by assuming $\tau_S \ll \tau_c $, and in the markovian regime $\tau_S \ll \tau_c \ll \tau_r$.  
As we are dealing with a structured environment, we shall start by  studying a weakly coupled system, which leads to a markovian regime  (i.e  $\gamma_0<\lambda$). Firstly, we shall compare a two level undriven ($\Delta=0$) evolution to an unitary one in order to see how different the open evolution is and decide whether the geometric phase can be measured in such scenario. Hence, in Fig. \ref{Fig4}, we show the total geometric phase accumulated for the non-unitary (red circled line) and unitary (blue asterisk line) evolution as time evolves,  being the number of cycles $N=\tau/\tau_S$. Therein, it is possible to see that initially the geometric phases are similar, with an estimate error of $2.5\%$ for 5 cycles and $10\%$ for 15 cycles when $\gamma_0=0.01$. As time evolves, the difference among both lines increases as expected, since for long times the loss of purity of the system would be considerable. 
  \begin{figure}[h]
	\includegraphics[width=9cm]{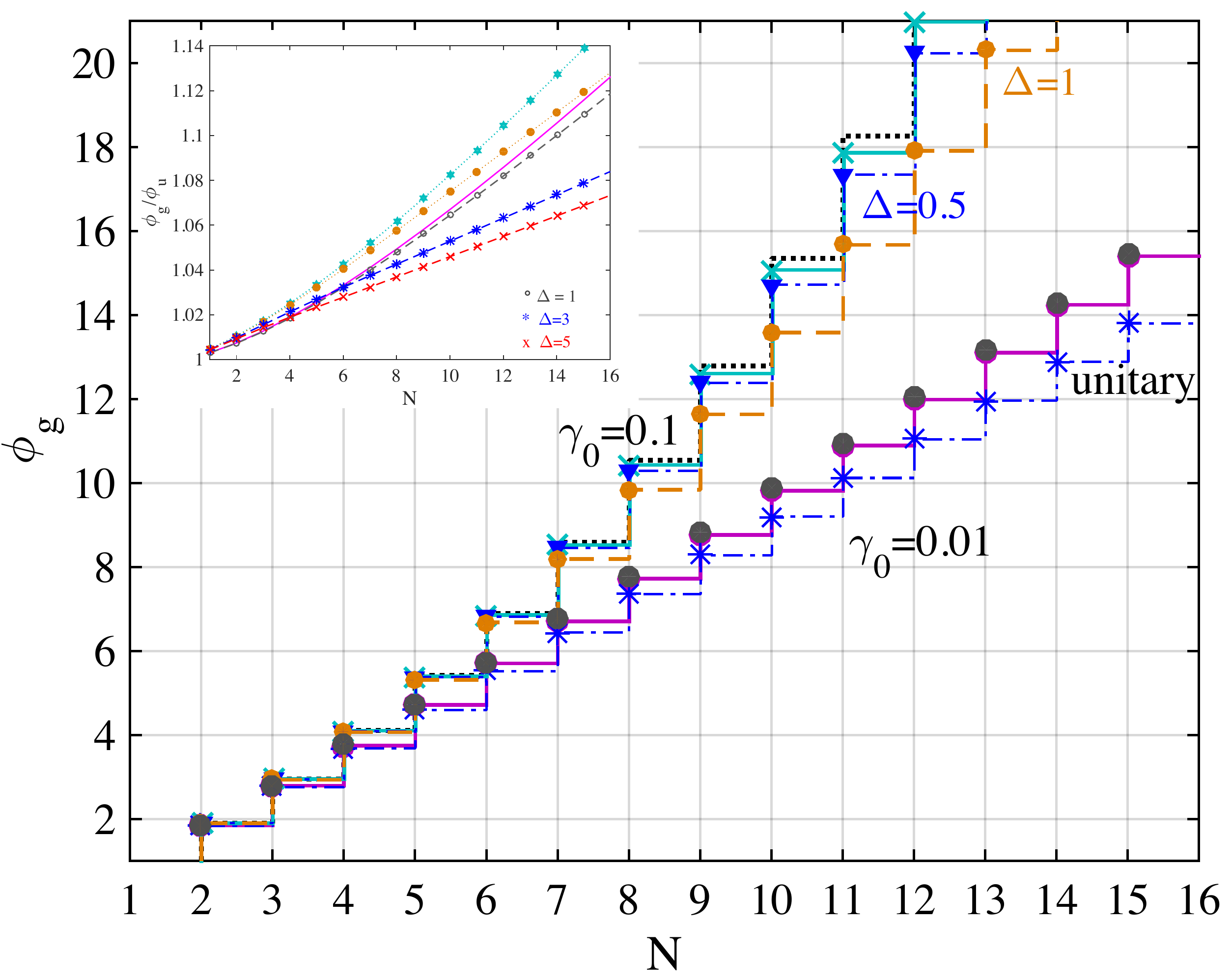}
 	\caption{(Color online) We study the effect of adding $\Delta$ by computing the geometric phase accumulated in time: $\Delta=0$ black dotted line, $\Delta=0.3$ turquoise cross line, $\Delta=0.5$ blue triangle line and $\Delta=1$ circle orange line for $\gamma_0=0.1$. For $\gamma_0=0.01$ gray squares ($\Delta=1$)  and  magenta solid line ($\Delta=0$) are very similar, while the blue asterisk line is the unitary geometric phase for reference. In the inset, we show the $\phi_g/\phi_u$ for $\gamma_0=0.01$ with different values of $\Delta$.  Parameters used:  $\Omega=20$, $\theta_0=\pi/4$ and $\omega_D=0$.}
 	\label{Fig5}
	 \end{figure}

In Fig. \ref{Fig5}, we show the geometric phase acquired when adding detuning frequencies to the two level system compared to the case when $\Delta \neq 0$, for different environments, say $\gamma_0=0.1$ and $\gamma_0=0.01$. When the coupling to the environment is very weak, the corrections to  the geometric phase acquired are very small and one can expect to obtain very similar results to the unitary geometric phase for few cycles of evolution. Evolutions with $\Delta \neq 0$ are very similar to $\Delta =0$  if we compared  the $\gamma_0=0.01$ results in Fig. \ref{Fig5} and Fig. \ref{Fig4}. In this case, the dominant correction to the geometric phase is given by the interaction with the environment, parametrized by the value of $\gamma_0$.
 However, for larger values of $\gamma_0$ ($\gamma_0=0.1$ but still weakly coupled), evolutions with bigger values of $\Delta$ acquire a geometric phase of considerable difference for long time evolutions. For the first few cycles, the geometric phases acquired are all very similar. As time evolves, different features as the magnitude of the coupling to the environment and the system's frequency (with $\Delta$ involved) have impact on the dynamics and therefore, in the geometric phase acquired.
 In the inset of Fig. \ref{Fig5}, we plot the normalized geometric phase ($\phi_g/\phi_u$) for $\gamma_0=0.01$.  We can see the $\Delta=0$ geometric phase represented by a magenta solid line, $\Delta=0.3$ turquoise cross line, $\Delta=0.5$ orange circled dotted  line,  $\Delta=1$ gray circled line, $\Delta=3$ blue asterisk line and $\Delta=5$ cross red line. The distance from unity becomes relevant as the number of cycles increases. 
 As expected, if $\Delta$ is added to the system, then the geometric phase acquired is different from that with $\Delta=0$, modifying the system's timescale involved and enhancing non-markovian effects as reported in \cite{Poggi}.  This can be a severe experimental problem to overcome. However, for  low-values of $\Delta$ considered here, the addition of a tunnel frequency does not considerably affect the geometric phase, obtaining  $\phi_g/\phi_u \sim 1$ for many evolution cycles in a weakly coupled regime.
 \begin{figure}[t]
	\includegraphics[width=9.cm]{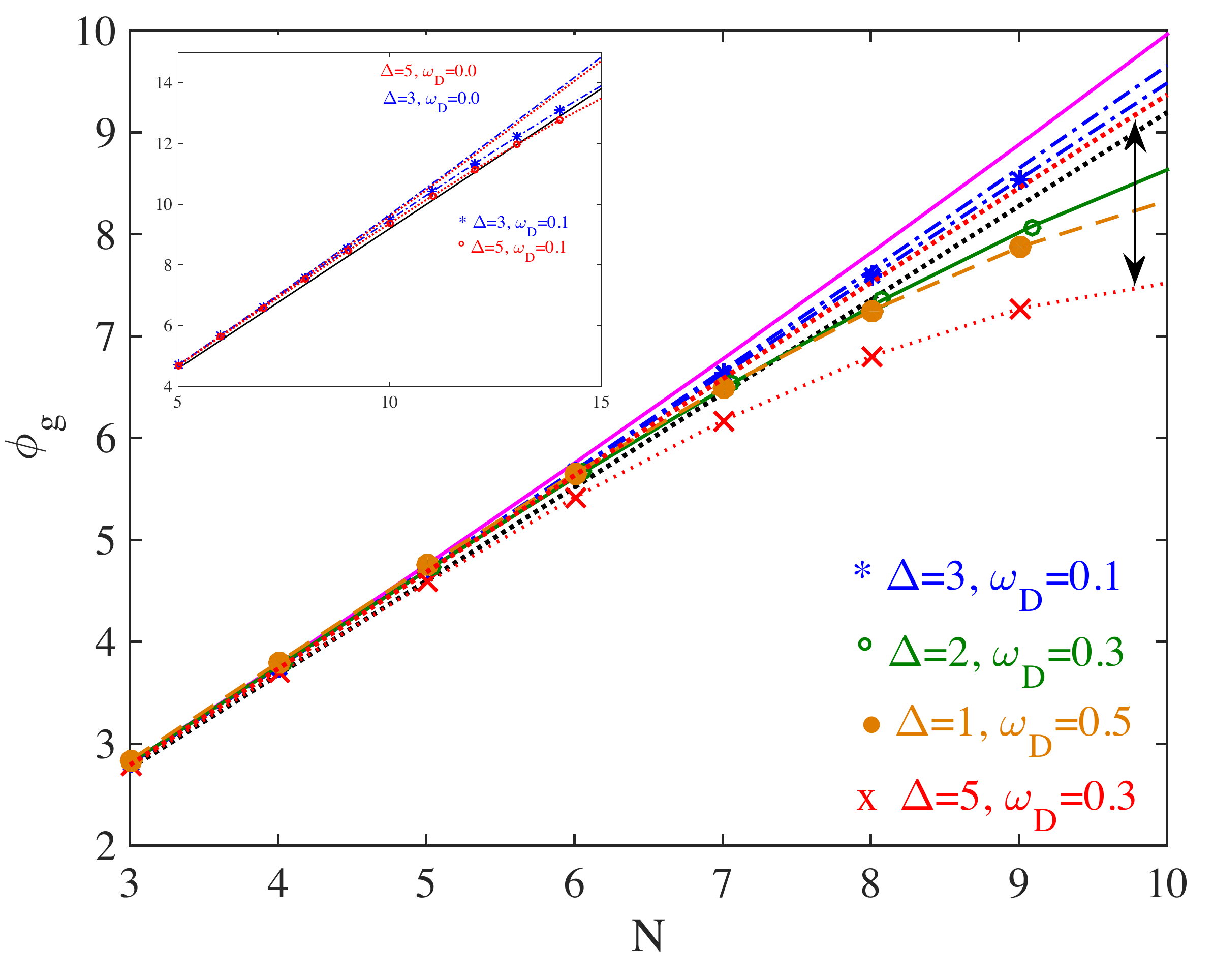}
 	\caption{(Color online) We include driving in the model and compute the geometric phase acquired $\phi_g$. Blue asterisk line correspond to $\omega_D=0.1$ and $\Delta=3$. Magenta line is for $\Delta=0$, red dotted line is for $\Delta=5$ and $\omega_D=0.3$, the green circled solid line for $\Delta=2$ and  $\omega_D=0.3$ and the orange circled dot-dashed line for $\Delta=1$ and  $\omega_D=0.5$. Black dotted unitary geometric phase is included for a reference. Low-frequency driving corrects the geometric phase accumulated for  short times. Parameters used: $\gamma_0=0.01$, $\Omega=20$, $\theta_0=\pi/4$.}
 	\label{Fig6}
 \end{figure}	 
 
We shall therefore study the interplay of adding driving to the two level system. In particular, we shall focus on the effect of driving when considering the possibility of measuring the geometric phase acquired by the two-state particle. In Fig. \ref{Fig6}, we show the geometric phase acquired when low-frequency driving is added: blue asterisk line correspond to $\Delta=3$ and $\omega_D=0.1$. Magenta line is for $\Delta=0$, red dotted line is for $\Delta=5$ and $\omega_D=0.3$, the green circled solid line for $\Delta=2$ and  $\omega_D=0.3$ and the orange circled dot-dashed line for $\Delta=1$ and  $\omega_D=0.5$. Black dotted unitary geometric phase is included for a reference. 
In the zoom plot we show the geometric phase acquired for $\Delta=3$ and $\omega_D=0$ (static) and compared it to $\Delta=3$ and $\omega_D=0.1$ (low-frequency field). We can see that the driven system acquires a geometric phase closer to the unitary one for longer periods of time, still the difference is very small.
In the main plot of Fig. \ref{Fig6}, we note that other driven systems are closer to the unitary geometric phase for ten periods as well. 
Therefore, there  are some set of parameters for which driving ``preserves purity". The geometric phase acquired is more similar to the unitary geometric phase acquired when there is  low frequency driving added for low values of $\Delta$. This fact can easily be observed in the inset plot, where the lines with asterisks and circles are closer to the unitary one (black solid line) than the corresponding static ones.
\begin{center}
\begin{figure}[t]
	\includegraphics[width=8cm]{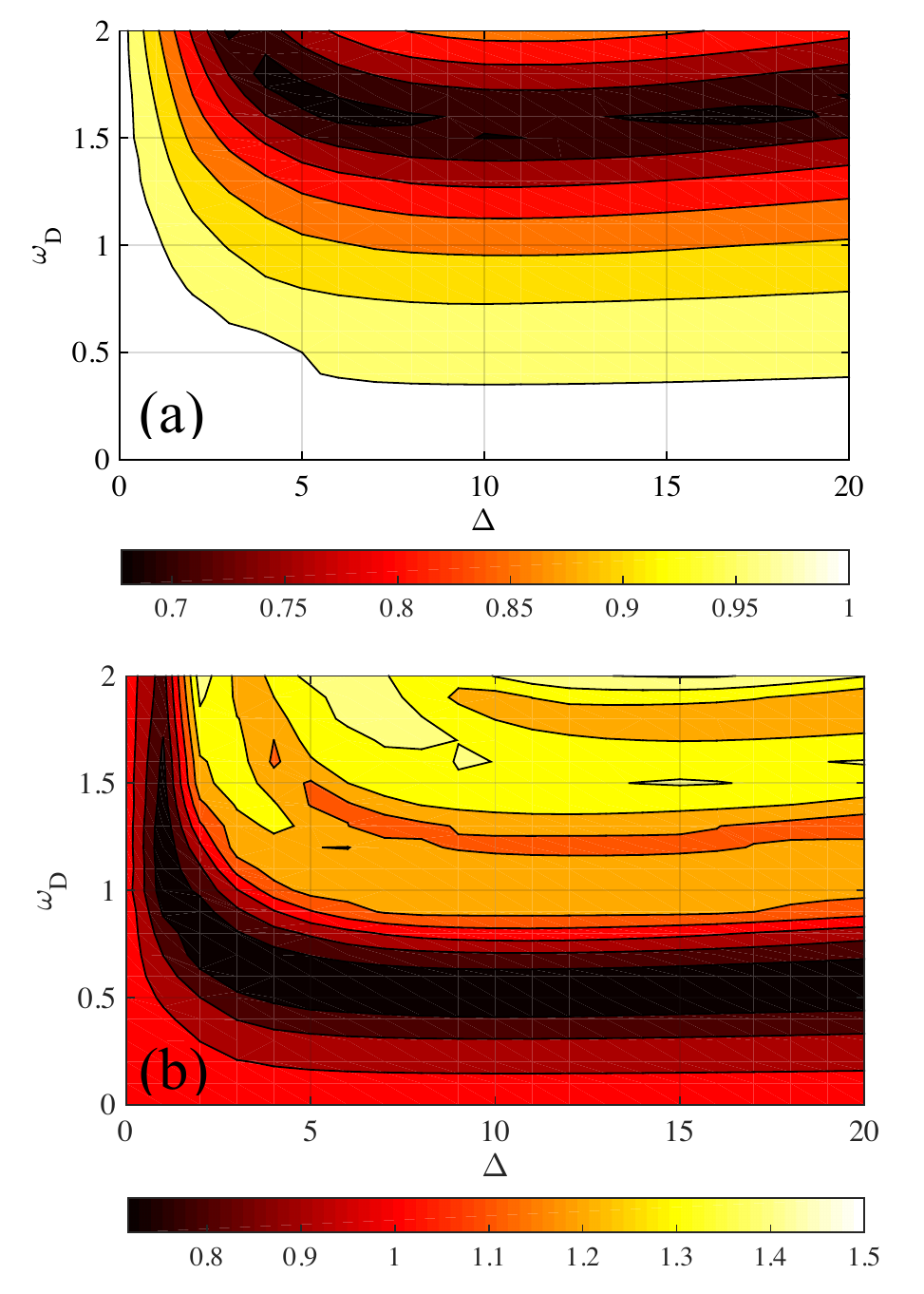}
 	\caption{(Color online)  $\phi_g/\phi_u$  for different values of $\Delta$ and $\omega_D$ for (a) $N=4$ and (b) $N=8$ under weakly coupling.  For short time evolution there is a wider set of parameters that yield $\phi_g/\phi_u \sim 1$. 
As time evolves, the set of parameters becomes smaller.
Parameters used: $\gamma_0=0.01$, $\Omega=20$, $\theta_0=\pi/4$.}
 	\label{Fig7}
\end{figure}
\end{center}
In Fig. \ref{Fig7}, we further explore this result by representing the normalized geometric phase $\phi_g/\phi_u$ as function of $\Delta$ and $\omega_D$, for two time evolutions: (a) $N=4$ and (b) $N=8$. It is easy to note that for short times, several model's parameters yield a $\phi_g/\phi_u \sim 1$. This can be understood because, as explained in Fig. \ref{Fig5}, the main contribution of the correction to geometric phase is given by the magnitude of the coupling between the environment and the system (say, if we assume $\phi_g=\phi_u + \delta \phi $ when $\gamma_0=0$, $\delta \phi=0$ and the geometric phase obtained is the unitary geometric phase $\phi_u$). However, as time evolves the intrinsic dynamic of each set of values ($\Delta, \omega_D$) will gain more importance. This type of behavior in the correction to the geometric phase has been observed in other studies, yielding that for short time the main correction derives from the fact that the environment is present and the system performs an ``open" evolution (and only markovian effects where taken into account) \cite{nature,ludmila}. 
As time elapses, the values of $\omega_D$ that preserve the unitary of the GP are less. For example, it can be seen in Figs. \ref{Fig6} and \ref{Fig7}, that $\Delta=5$ and $\omega_D=0.1$ renders a value $\phi_g$ closer to $\phi_u$ than $\Delta=5$ and $\omega_D=0.3$.
Likewise, adding a very low frequency driving and small detuning frequencies for short time evolutions renders a geometric phase similar to the unitary geometric phase, which leads to a good scenario of measuring the geometric phase in structured environments. 
It has been shown in \cite{Poggi} that $\omega_D/\Delta <1$ increases the degree of non markovianity (for a small coupling), and particularly, non markovianity increases with $\gamma_0$ and decreases with $\Delta$ for a given environment (fixed $\gamma_0$).  We are not strictly in the regime reported in \cite{Poggi}, since we are studying the situation for different evolving times. 
However, we must say that if we want to maintain the markovian regime, we should only add low detuning frequencies (if any at all), because by adding detuning frequencies  and driving frequencies we shall be modifying considerably the dynamics of the system and comparison with the undriven situation will be useless. However, we can still note that when adding a low-frequency driving, geometric phases acquired are very similar to the non-driven isolated geometric phases for bigger values of $\Delta$. This fact agrees with the result obtained in \cite{lofranco}, where authors state that for a small qubit classical field coupling, a non-resonant control ($\Delta \neq 0 $) is more convenient to stabilize the geometric phase of the open qubit. The use of driven systems can help the measurement of geometric phases under some set of parameters. This knowledge  can aid the search for physical set-ups that best retain quantum properties under dissipative dynamics.  As can be inferred for the different simulations done, for weak coupling, the better scenario would be to have a small detuning frequency and very low frequency driving field, so as to maintain the smoothness of a markovian evolution and acquire a geometric phase similar to the unitary geometric phase.
 
\begin{figure*}[t]
 	\includegraphics[width=16.cm]{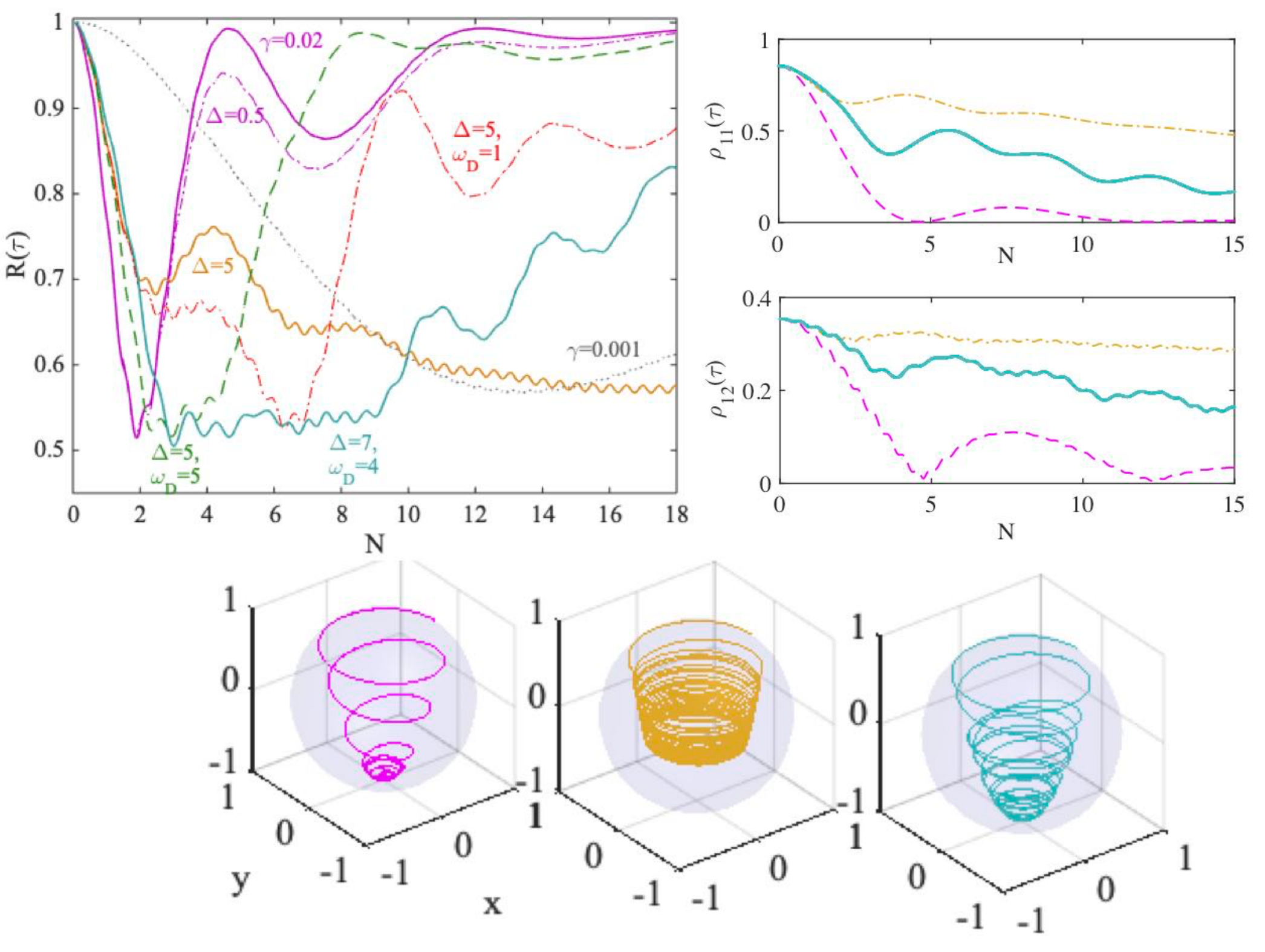}
 	\caption{(Color online) Top: Bloch vector ($R(\tau)$) temporal evolution  for different set of model parameters. On the right corner we show some matrix elements:  population  $\rho_{11}(\tau)$ and the absolute value of the off-diagonal term $\rho_{12}(\tau)$  for three different set of parameters.
	Bottom: Trajectories in the Bloch sphere ($\vec{R}=(x,y,z)$) for three sets of parameters with $\gamma_0/\lambda \geq 1 $: in magenta, $\omega_D=0=\Delta$; orange for  $\omega_D=0$ and $\Delta=5$ and light-blue for $\omega_D=4$ and $\Delta=7$. In all cases $\tau_c=100$ and $\Omega=20$.}
 	\label{Fig8}
\end{figure*}

\subsection{Geometric phase under an intermediate coupling}

\begin{figure}[h!]
	\includegraphics[width=9.1cm]{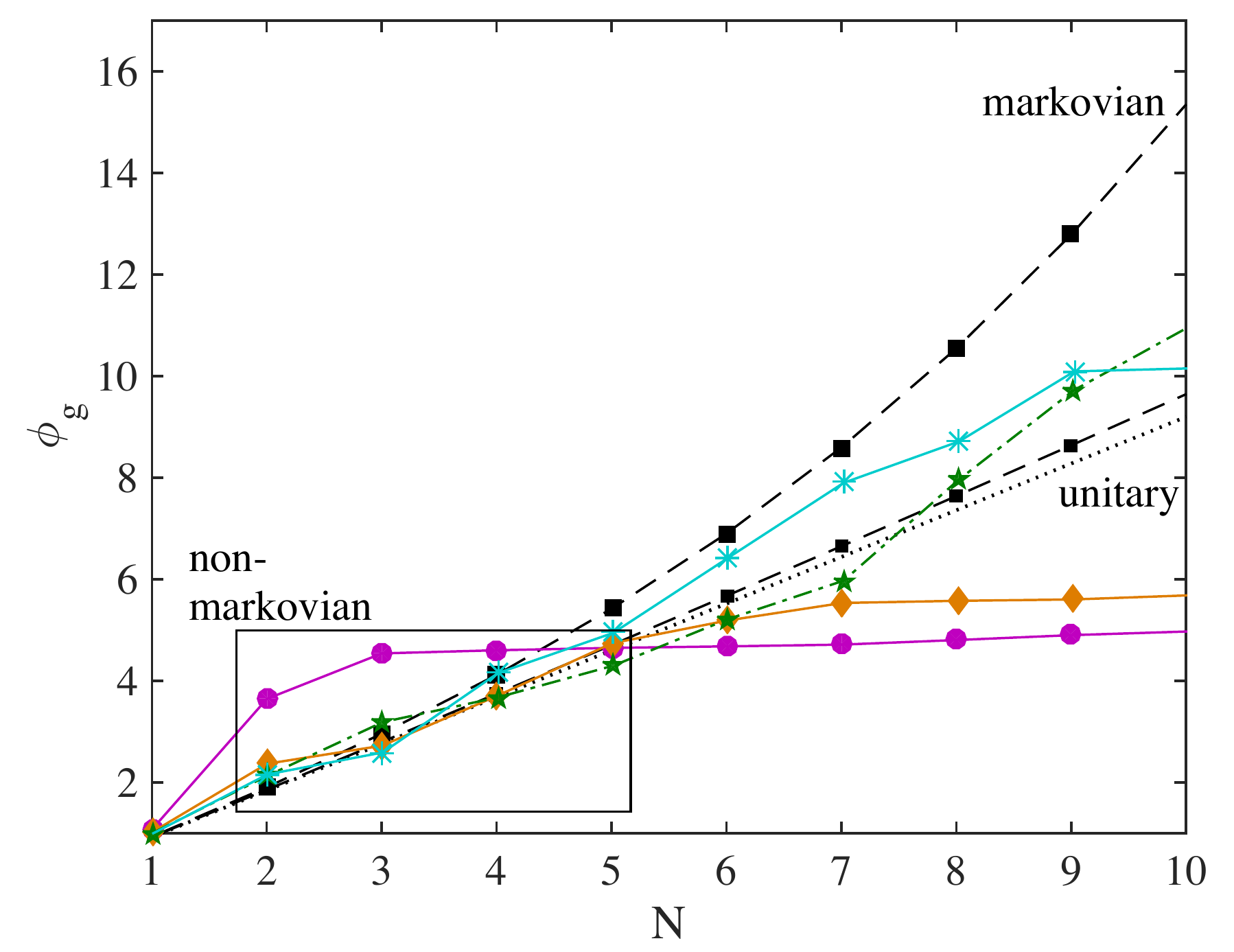}
 	\caption{(Color online) Geometric phase accumulated $\phi_g$ as number of periods evolved for different set of parameters. 
	Colors represent the parameters:    black dotted line is the unitary geometric phase; the black squared line if for a markovian evolution as described above; in magenta non-markovian evolution for  $\omega_D=0=\Delta$; dot-dashed diamond orange for  $\Delta=5$  and $\omega_D=5$;  light-blue asterisk line for $\Delta=7$ and $\omega_D=4$. Parameters used: $\gamma_0=1$, $\Omega=20$, $\theta_0=\pi/4$.}
 	\label{Fig9}
\end{figure}
In the above section we showed the geometric phase acquired by the two-level driven system in a weakly coupled structured environment. 
The above selection of parameters rendered a markovian situation where one could still find evidence of a quasi-cyclic evolution, since the degradation of the pure state was done slowly and there were no revivals. 
 In the following, we shall show what happens if the parameters are chosen so as to simulate a 
 non-markovian environment by considering $\bar{\gamma_0}/\lambda> 0.25$, as it  has been said in \cite{Poggi}.
This situation can model for example a  two-level emitter  with a transition frequency driven by an external classical field of frequency $\omega_D$ embedded in a zero temperature reservoir formed by the quantized modes of a high-Q cavity.
In such a case, the evolution is wildly modified and one can find revivals after a given number of periods.
This shall help us to understand the role of driving in this type of environments. If we set parameters so as to see non-markovian behavior, then we must mention that finding tracks of the geometric phase can be much more difficult. In  Fig. \ref{Fig8}, we show  different scenarios by setting the model parameters. 
On top of Fig. \ref{Fig8}, we show the temporal evolution of the Bloch vector ($|R(\tau)|$) for different driven frequencies with $\bar{\gamma_0}$ and fixed $\lambda$.   As can be seen, this type of environment starts to exhibit non-markovian environment though revivals are small in amplitude: the magenta line represents $\Delta=0.0=\omega_D$;
the dot-dashed magenta line is for  $\Delta=0.5$ and $\omega_D=0$. The orange line is for $\Delta=5$; while the red dot-dashed line is for $\Delta=5$ but $\omega_D=0.1$. The dotted green line  represents $\Delta=5$ and $\omega_D=5$  and dot-dashed cyan line  $\Delta=7$ and $\omega_D=4$. We have also included a markovian evolution just for reference (black dotted line for $\gamma_0=0.001$).
We can easily note that the amount of driving changes considerably the evolution of the initial quantum state. On the top right corner we show the populations probability for different lines: magenta dashed  line, represents the $\Delta=0.0=\omega_D$, the orange dotted lined $\Delta=5$, and the cyan dashed line  $\omega_D=4$ and $\Delta=7$. We can see that by adding a frequency $\Delta$ and a driving frequency $\omega_D$, revivals disappear, recuperating the opportunity to track traces of a geometric phase. This fact can be easily observed in the Bloch sphere.
At the bottom of Fig. \ref{Fig8}, we represent the trajectory ($\vec{R}=(x(\tau),y(\tau),z(\tau)$) in the Bloch sphere of the initial state of the three different sets of parameters for the same number of cycles evolved. We can see that the transition among states is done in a short time for the magenta line. The revivals stimulate the exploration of the south pole of the Bloch sphere, for another period of time until it finally decays. In such an evolution, one can only achieve a geometric phase during the revivals and compare it to the one the system would have acquired if it has started at that latitude of the Bloch sphere. In the case of the orange line, transition among states is delayed by the frequency change of the system's period $\tau_S=2 \pi/(\Omega+\Delta)$. In this case, the geometric phase can be measured for very short initial periods. Finally, for the cyan curve we can observe that the evolution remains ``frozen" at a latitude for almost 3 cycles before continuing the transition among states.
\begin{figure}[h!]
 	\includegraphics[width=9.1cm]{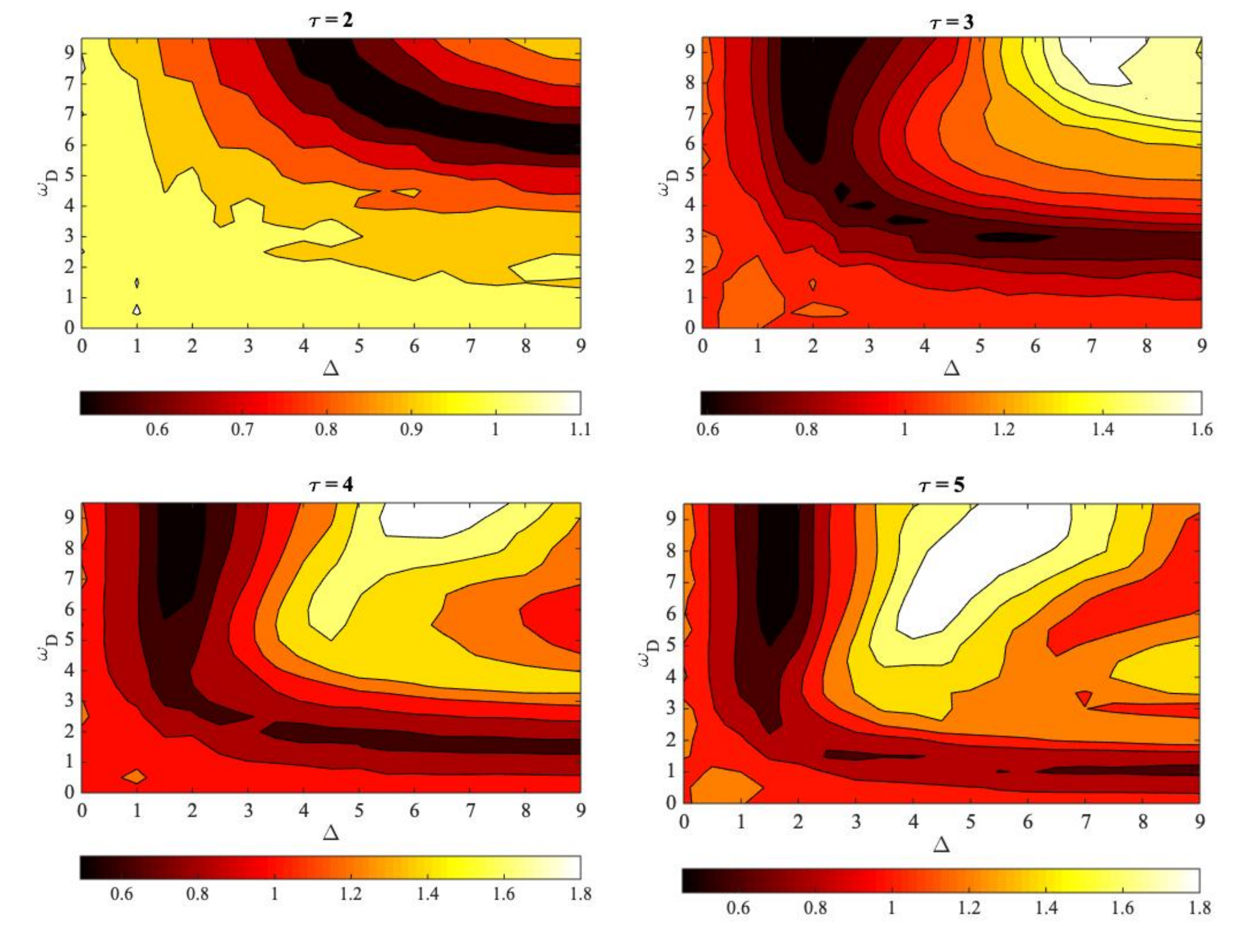}
 	\caption{(Color online) Geometric phase accumulated $\phi_t/\phi_g$ in the ($\Delta$, $\omega_D$) plane for different number of periods evolved  in a non markovian environment: $N=2$, $N=3$, $N=4$ and $N=5$. Parameters used: $\gamma_0=1$, $\Omega=20$, $\theta_0=\pi/4$. }
 	\label{Fig10}
\end{figure}

We can therefore compute the geometric phase for these different situations in order to see if it is possible to track traces of an accumulated geometric phase during the evolutions. In Fig. \ref{Fig9} we show the geometric phase accumulated for different set of parameters  (as  those considered in Fig. \ref{Fig8}). The colors of the lines in Fig. \ref{Fig9} correspond to the same values of Fig.  \ref{Fig8}. The magenta line (with dots) is the temporal evolution of an initial state  under a structured environment in a non-markovian regime with $\Delta=0$. In this case after 4 periods, the evolution presents some revivals after having made a transition from the upper to the lower state (therefore revivals are done in the south pole sphere). This is easily understood with the information given in Fig. \ref{Fig8} where we see that transition is done at very short times. Therefore,  the geometric phase acquired for $\Delta=0$ is very different to that the system would have acquired in a markovian regime (black squared solid line) or an isolated evolution (black  dotted line). However, in Fig. \ref{Fig9}, we also present the geometric phase for driven systems under non-markovian regime. The diamond orange line represents a driven case of $\omega_D=5$ and $\Delta=5$. In such situation, we see that the evolution of the system initially recovers some ``unitarity", acquiring a geometric phase very similar to that of the unitary case. Finally,  after some periods, it makes a transition and the evolution explores the south pole sphere. Finally, the light-blue asterisk line for $\omega_D=4$ and $\Delta=7$, acquires a geometric phase similar to the markovian one for longer time periods.  In this last driven case, we see that adding driving has a relevant consequence: the geometric phase acquired is closer to the one acquired under a markovian evolution, and therefore closer to the unitary one for a short number of periods evolved ($\sim N=10$). For smaller time periods, we see that adding driving preserves the geometric phase: in all cases showed, the geometric phase is recovered compared to the case when $\Delta=0$.
Finally, in Fig. \ref{Fig10} we show a general scenario of the situation described above for $\phi_g/\phi_u$ at different times: $N=2$, $N=3$, $N=4$ and $N=5$. We effectively notice regions of the $\omega_D-\Delta$ space where the accumulated geometric phase $\phi_g/\phi_u$ remains close to one, meaning that the geometric phase acquired is close to the unitary one. 
There are regions where $\phi_g$ departs enormously from $\phi_u$.
As this situation exhibits non-markovian effects as revivals, it is not that easy to find a general rule so as to when it is more convenient to measure the geometric phase. 
However, we must say that there are some situations where driving enhances the ``robustness" condition of the geometric phase when $\Delta$ delays the revivals and $\omega_D$ is small.
We can see that for some particular situations, the addition of a frequency $\Delta$ and driving $\omega_D$ becomes a useful scenario so as to get control of the geometric phase. These situations deal with a smoothening of the revivals as shown in Fig. \ref{Fig8} for the cyan curve. In \cite{Poggi}, authors show that there is a large region, corresponding to $\omega_D/\Delta \sim  {\cal O}(1)$ where non-markovianity is suppressed altogether for an intermediate coupling.  They even state that in a strong coupling regime ($\gamma_0>1$), the driving is unable to increase the degree of non-markovianity,  contrary to what one can expect when adding driving to the system. On this aspect, authors in \cite{lofranco} state that intense classical fields strongly reduce non-markovianity of the system. To prevent this, they state that the larger the coupling, the higher the values of detuning required in order to maintain a given degree of non-markovianity when dealing with hybrid quantum-classical systems. 
Herein, we assume an intermediate coupling $\gamma_0=1$, where there are some parameters $\omega_D/\Delta \sim  {\cal O}(1)$ that verify a suppression of revivals and assure a smooth evolution and an acquisition of a geometric phase more similar to the unitary one. 
This fact, in addition to some other features related to the initial quantum state explained below, contribute to a better understanding of driven systems and should be taken into consideration when designing experimental set-ups to measure geometric phases.

\subsubsection{Dependence on $\rho_s(0)$}

In this section we shall study the dependence upon the initial state of the quantum system. As explained above, we consider an initial pure state of the form $
|\Psi (0) \rangle=\cos(\theta_0/2) |0 \rangle
+ \sin(\theta_0/2) |1 \rangle,$
with $0\leq \theta_0 \leq \pi/2  $. 
This determines the initial values of the reduced density matrix $\rho_{11}(0)=\cos(\theta_0)^2$ and $\rho_{12}(0)=1/2 \sin(2 \theta_0)$.  In the manuscript, we have always started with an initial $\theta_0=\pi/4$, so as to consider an initial average state (where the geometric phase is more ``stable"). 
In the following, we shall study how decoherence affects different initial states of the two-level system.
We shall use  the change in time of the absolute value of $|R(\tau)|=R(\tau)$ as a measure of decoherence.
 In Fig. \ref{fig1ap}, we show $R(\tau)$ as function of time: (a) is for a weakly coupling  while (b) for an intermediate coupling. The curves start from an angle of $\theta_0=0.15$ radians (near the north pole of the Bloch sphere) to $\theta_0=1.5$ radians (near the Equator). The measurement of the ``robustness"  of quantum states is that the loss of purity of the state vector is very small for many cycles. The dependence of this magnitude upon time (measured in cycles) depends on the initial quantum state for the same parameters of the model. As it can be seen there, the state is more affected for smaller initial angles. The purity of the state remains close to unity (isolated case) when the initial state is located near the equator of the Bloch Sphere ($\theta_0 \sim \pi/2$) for both environments. This  means an initial state of the form $|+\rangle=\cos(\pi/4)|0\rangle + \sin(\pi/4)|1\rangle$. This can be understand by noting that the Interaction Hamiltonian is proportional to $\sigma_x$, which in turn is  eigenstate of the Interaction Hamiltonian. 

As for the experimental detection of the geometric phase, we need to find a compromise between the loss of purity and the area enclosed in the path trajectory. As the natural evolution of the system would be to make a transition to the lower state of the quantum state, we need to ``control" this evolution so as to obtain small variations of  the trajectory and still find traces of the geometric phase. The non-markovian evolution is the one that provides the more interesting results and the one that can be used to model experimental situations such as hybrid quantum classical systems feasible with current technologies, so we shall explore in detail the dependence upon the initial angles.

\begin{widetext}
\begin{center}
\begin{figure}[h]
 	\includegraphics[width=17cm]{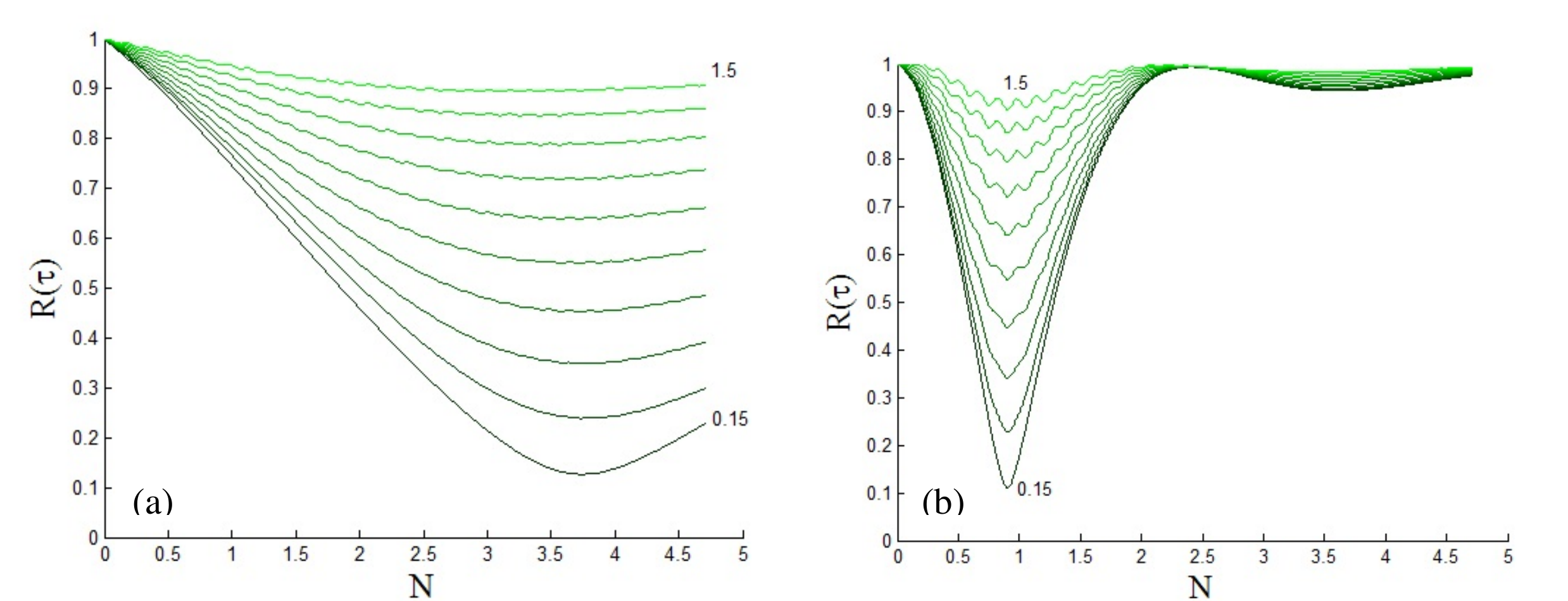}
 	\caption{(Color online)  Loss of purity $R(\tau)$ as function of time for (a) a weakly coupling ($\gamma_0=0.01$) and (b) an intermediate coupling ($\gamma_0=1$) for different initial angles $\theta_0$. We can see that as $\theta_0$ reaches $\pi/2$ (labeled as 1.5) the curves are closer to the unitary evolution.}
 	\label{fig1ap}
\end{figure}
\end{center}
\end{widetext}

\begin{figure}[h!]
 	\includegraphics[width=9.1cm]{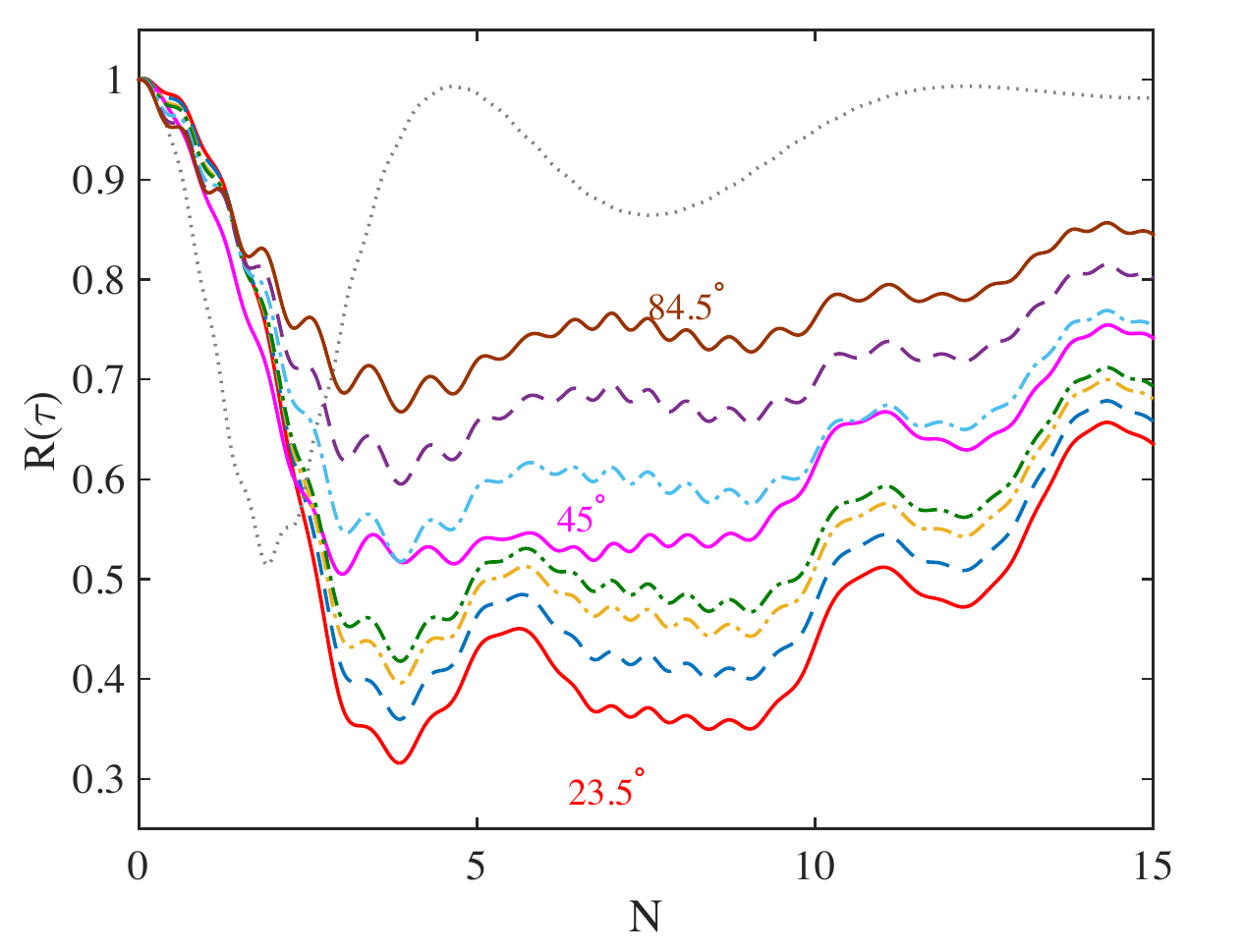}
 	\caption{(Color online) $R(\tau)$ as function of the number of cycles evolved for a non-markovian environment with driving for different values of the initial quantum state ($\theta_0$ labeled in degrees). For reference, we also included the static non-markovian evolution $\omega_D=0=\Delta$ with a black dotted line.	Parameters used: $\gamma_0=1$, $\Omega=20$, $\omega_D=4$ and $\Delta=7$.}
 	\label{fig2ap}
\end{figure}

In Fig. \ref{fig2ap}, we show the loss of purity $R(\tau)$ for a non-markovian evolution with driving considered in the manuscript (included in Fig. \ref{Fig9}) for different initial angles $\theta_0$: the lower initial angle considered is $23.5^\circ$ with a red solid line and a brown solid line for the bigger angle considered $\theta_0=84.5^\circ$. In between, we have considered several angles $\theta_0=35^\circ$, $\theta_0=40^\circ$, $\theta_0=45^\circ$, $\theta_0=50^\circ$, $\theta_0=62^\circ$ and $\theta_0=73^\circ$. For reference, we also included the static non-markovian evolution $\omega_D=0=\Delta$ with a black dotted line.
\begin{figure}[h!]
 	\includegraphics[width=9.1cm]{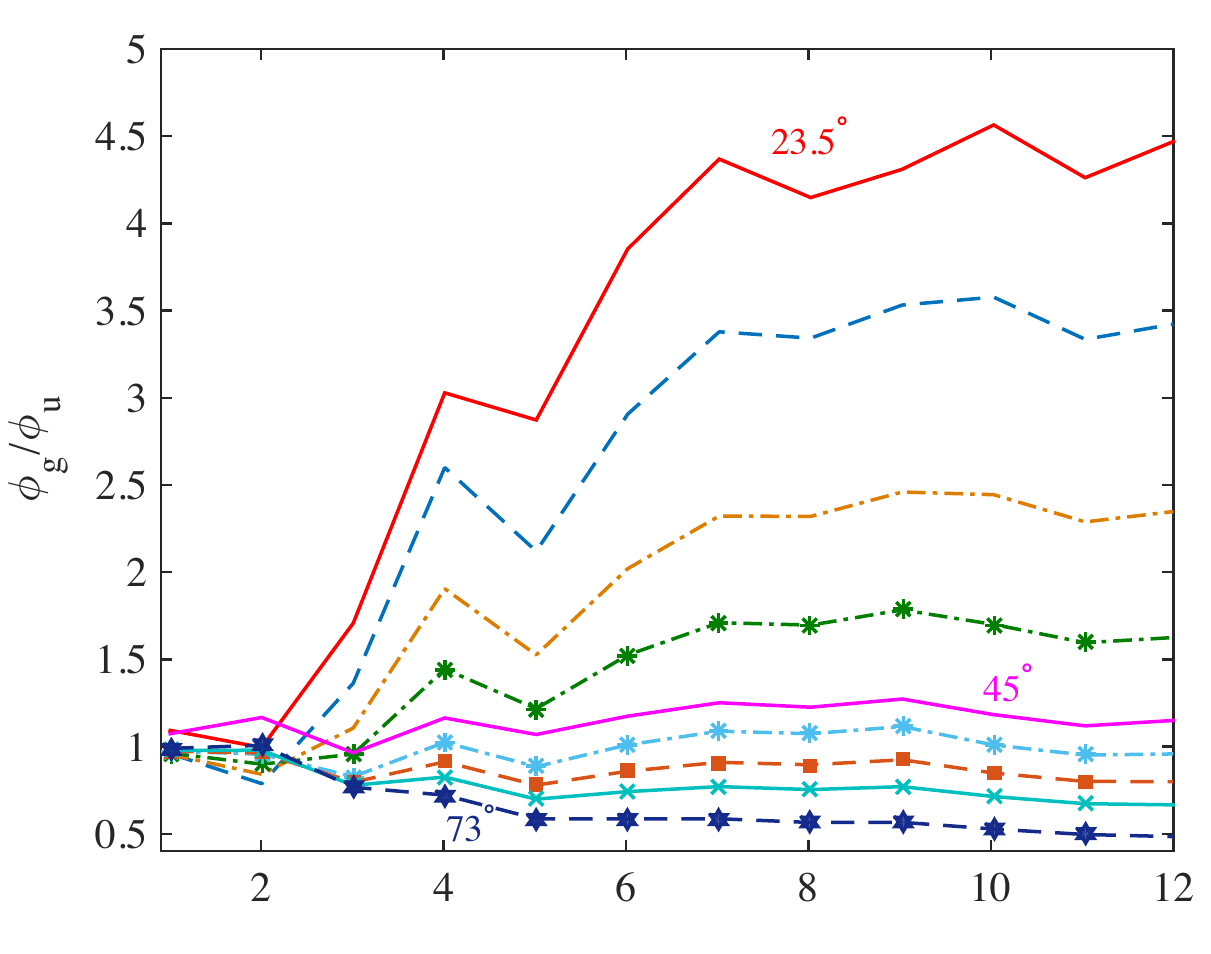}
 	\caption{(Color online) Geometric phase accumulated normalized  ($\phi_g/\phi_u$) by the unitary geometric phase accumulated for several cycles of the evolution under a non-markovian environment for different initial angles.  Parameters used:  $\gamma_0=1$, $\Omega=20$, $\omega_D=4$ and $\Delta=7$.}
 	\label{fig3ap}
\end{figure}

We can see that all cases considered are qualitatively similar, however $\theta_0 \sim \pi/4$ is the one that maintains the degree of purity for several cycles. This fact can be fruitfully exploited for the detection of the geometric phase. In Fig. \ref{fig3ap}, we show the geometric phase accumulated normalized by the unitary geometric phase accumulated for several cycles of the evolution under a non-markovian environment for different initial angles. The lower angle considered is $23.5^\circ$ and we can see that the GP acquired is very different to the unitary one ($\phi_g/\phi_u \neq 1$). We considered increasing initial angles up to $45^\circ$ indicated by a magenta solid lined which gives a $\phi_g/\phi_u $ close to 1 for several cycles (in agreement to results shown in  Fig. \ref{Fig9}). The next light blue-asterisk line is for $\theta_0=50^\circ$ and shows a similar behavior. Angles continue to increase up to the hexagram blue curve indicating $\theta_0=73^\circ$. Therein, we can see that as the angle increases the difference between $\phi_g$ and $\phi_u$ grows becoming considerable for large angles. That is the reason we believe that in order to experimentally detect the geometric phase we need to only consider the decoherence model of noise but the geometric aspects of $SU(2)$ as well.

\section{Conclusions}
\label{conclusiones}
In this manuscript, we have focused on the hierarchy equations of motion  method in order to study the interplay between driving and geometric phases.  This method can be used if (i) the initial state of the system plus the bath is separable and (ii) the interaction Hamiltonian is bilinear. It results in an advantageous method since it provides a tool to simulate markovian and non markovian behavior in the structured spectrum. 

We have therefore studied the dynamics of the system and computed the geometric phase for different environment regimes defined by the relation among the model's parameters. In all cases we have focused on the effect of adding driving to the two-state system.
By numerically studying the proposed model for various parameter regimes, we find a remarkable result: the driving can produce a large enhancement of non markovian effects, but only when the coupling between system and environment is small.
 We have seen for a weakly coupled configuration, 
 when adding  a low frequency driving to the quantum system's frequency, the system's dynamic tends to be corrected towards the undriven situation only for very small values of $\omega_D$. This can be understood that by adding a detuning frequency changes considerably the system's timescale and therefore, the geometric phase would be different from the unitary undriven one. 
 More interesting, for a stronger coupling or non markovian regime, that there are some situations where driving enhances the ``robustness" condition of the geometric phase when $\Delta$ delays the revivals and $\omega_D$ is small, particularly when $\omega_D/\Delta \sim {\cal O}(1)$. 
As stated in the existing bibliography, in  strong-coupling regime, on the other hand, the driving is unable to increase the degree of non markovianity. In this manuscript, we have further studied the intermediate coupling, since we try to track traces of the geometric phase, which is literally destroyed under a strong influence of the environment. In this regime, where there are some non markovian features characterizing the dynamics, we have noted a suppression of non markovianity altogether, allowing for a smooth dynamical evolution. We have further shown
that for low-frequency driving, the driving fails to increase the degree of non markovianity with respect to the static case, recuperating in some cases a scenario where a geometric phase can still be measured ($\phi_g=\phi_u +\delta \phi$).  This knowledge can aid the search for physical set-ups that best retain quantum properties under dissipative dynamics.

As we have noticed that the non markovian evolution (with intermediate coupling) is the situation that provides the more interesting results, and further it is the one that can be used to model experimental situations such as hybrid quantum classical systems feasible with current technologies, we have explored in detail the dependence upon the initial angles for a better understanding of the results. We have found that exits a set of  more ``stable" initial angles.
This means that while there are dissipative and diffusive effects that induce a correction to the unitary GP, the system maintains its purity for
several cycles, which allows the GP to be observed. It is important to note that if the noise effects induced on the system are of considerable magnitude, the coherence
terms of the quantum system are rapidly destroyed and the GP literally disappears.
 It has been argued
that the observation of GPs should be done in times long enough to obey the adiabatic approximation but short enough to prevent decoherence from deleting all
phase information. As the geometric phase accumulates over time, its correction becomes relevant at a relative short timescale, while the system still preserves purity.
All the above considerations lead to a scenario where the geometric phase can still be found and it can help us infer features of the quantum system that otherwise might be hidden to us. 
 \\

We acknowledge UBA, CONICET and ANPCyT--Argentina.   
The authors wish to express their gratitude to the TUPAC cluster, where the calculations of this paper have been carried out.
We thank F. Lombardo and P. Poggi for their warmhearted discussions and comments.\\

{}

\end{document}